\documentstyle[twocolumn]{article}

\renewcommand{\H}{I\!\!H}

\begin{document}
\title{Mode decomposition and unitarity in quantum cosmology
        \thanks{Talk given at the {\it Second Meeting on Constrained
 Dynamics and Quantum gravity}, Santa Margherita Ligure, 
 September 17--21, 1996.}}
 
\author{Franz Embacher\\ \smallskip 
       Institut f{\"u}r Theoretische Physik, 
        Universit\"at Wien,\\ Boltzmanngasse 5, A-1090 Wien\\
        E-mail: fe@pap.univie.ac.at\\
        \\
        UWThPh-1996-67\\
        gr-qc/9611055 }

\date{} 
\maketitle

\begin{abstract}
Contrary to common belief, there are perspectives for
generalizing the notion of positive and negative frequency 
in minisuperspace quantum cosmology, even when the wave equation
does not admit symmetries. We outline a strategy in doing so
when the potential is positive. Also, an underlying unitarity
structure shows up. Starting in the framework of the Klein-Gordon 
type quantization, I am led to a result that relies on global
features on the model, and that is possibly related to structures
encountered in the refined algebraic quantization scheme. 
\end{abstract}

\section{INTRODUCTION}

The basic ingredients of minisuperspace quantum cosmology are
an $n$-dimensional (minisuperspace) manifold $\cal M$, 
endowed with a metric 
$ds^2 =g_{\alpha\beta}\,dy^\alpha dy^\beta$ 
of Lorentzian signature $(-,+,\dots,+)$
and a potential function 
$U$ (which is assumed to be positive).  
The wave equation (the minisuperspace version of the
Wheeler-DeWitt equation) reads   
\begin{equation}
\Big(-\nabla_\alpha \nabla^\alpha + U \Big)\,\psi = 0.   
\label{2.2}
\end{equation}
There are several mathematical structures associated with 
these ingredients, and accordingly one may take several
routes when ''solving'' this equation, giving the solutions 
some sense as ''quantum states'' of the (mini)universe and
making contact to the mathematical structures of conventional 
quantum mechanics. 

There are two condidate scalar products available, 
\begin{equation}
q(\psi_1,\psi_2)=\int_{\cal M} d^n y\,\sqrt{-g}\,\psi_1^*\psi_2 
\label{q}
\end{equation} 
and
\begin{equation}
Q(\psi_1,\psi_2)= -\,\frac{i}{2}\, \int_\Sigma d\Sigma^\alpha\,
(\psi_1^*\stackrel{\leftrightarrow}{\nabla}_\alpha \psi_2) \, , 
\label{2.8}
\end{equation}
where $\Sigma$ is a spacelike 
hypersurface (of sufficiently regular asymptotic behaviour, 
if necessary). The old-fashioned Klein-Gordon quantization
\cite{Kuchar} 
is based on the observation that the indefinite 
scalar product $Q$ is independent of $\Sigma$ if
both functions $\psi_{1,2}$ satisfy the wave equation. It 
benefits from the mathematical similarity of the 
quantum cosmological framework 
to the quantization of a scalar particle in a curved space-time
background $({\cal M},ds^2)$ and an external positive potential $U$. 
One of the key paradigms along this road is that --- in the 
generic case --- it is not 
possible to decompose the space $\H$ of wave functions into two 
subspaces $\H^\pm$ consisting of positive and negative frequency 
modes \cite{Kuchar} (within which $Q$ would
be positive and negative definite, respectively). 
A possible reaction to this fact is the point of view that the
mathematics of quantum mechanics (in particular a positive definite
scalar product) emerges only at an approximate level in a 
semiclassical context. 
Other routes to quantization start from the positive
definite scalar product $q$, in particular the refined algebraic
quantization \cite{ref}, 
regarding $(L^2({\cal M}),d^n y\,\sqrt{-g}\,)$ as an
auxiliary Hilbert space from which the space of physical states
is constructued by distributional techniques. 

Here, I would like to begin with the conservative point of view of
Klein-Gordon quantization and ask whether a preferred decomposition 
$\H=\H^+\oplus\H^-$ of the space of wave functions may be constructed. 
I will {\it not} attempt to define positive and negative frequencies 
(in the sense of $\psi^\pm\sim e^{\mp i\omega t}$ with respect to 
some time coordinate) but try to {\it generalize} 
this concept. During the procedure, I will be led towards
non-local features in a quite natural way. The strategy 
is rather formal, and the class of models to which it
may be applied is not very well known. However, it gives a hint
how different quantization methods, relying on the two
scalar products $q$ and $Q$, might be related in terms of some 
underlying structure. The details of the computations outlined 
here may be found in Ref. \cite{FE10}. 

\section{UNITARITY AND PREFERRED DECOMPOSITION}

I will use a congruence of hypersurface orthogonal classical 
trajectories 
as background structure with respect to which the wave equation
is further analyzed. In what follows I will assume that all
quantities are defined in domains large enough to allow for  
the existence of the integrals appearing below. 
Let $S$ be a solution of the Hamilton-Jacobi equation 
\begin{equation}
(\nabla_\alpha S) (\nabla^\alpha S)= -\, U 
\label{2.5}
\end{equation}
and $D$ a positive solution 
of the conservation equation 
\begin{equation}
\nabla_\alpha (D^2 \nabla^\alpha S) = 0\, . 
\label{3.6}
\end{equation}
The action function $S$ generates a congruence of classical
trajectories by means of the differential equation 
\begin{equation}
\frac{1}{N} \, \frac{dy^\alpha}{dt} = 
\nabla^\alpha S \, , 
\label{2.6}
\end{equation}
with $N$ an arbitrary lapse function, 
and $D$ provides a weight on the set of trajectories, thus 
specifying the scalar product defined in (\ref{3.13}) below. 
The pair $(D,S)$ is called a WKB-branch (although I will not
impose any approximations here). 
The (''time'') evolution parameter is defined by $t=-S$, and 
one may choose coordinates $\xi^a$ labelling the 
trajectories. This corresponds to the lapse 
funcion $N=U^{-1}$ in (\ref{2.6}), and the metric becomes 
\begin{equation}
ds^2 = -\,\frac{dt^2}{U} + \gamma_{ab}\,d\xi^a d\xi^b\,. 
\label{3.3}
\end{equation}
By $\partial_t$ (or a dot) I denote the derivative along the
trajectories (in the coordinates $(t,\xi^a)$ it is just the partial 
derivative with respect to $t$). 

Given a WKB-branch, any wave function may be written as
\begin{equation}
\psi = \chi D e^{i S} \, , 
\label{3.8}
\end{equation}
by which the wave equation (\ref{2.2}) becomes 
\begin{equation}
i\, \partial_t \chi = \Big(\,\frac{1}{2}\,\partial_{tt} + h \Big)
\chi \, , 
\label{3.9}
\end{equation}
where 
\begin{equation}
h = H^{\rm eff} - \frac{1}{2}  \,D \,(D^{-1})\,\dot{}\,\,\dot{}
\label{3.10}
\end{equation}
and 
\begin{equation}
H^{\rm eff} = 
-\,\frac{1}{2 D \sqrt{U}}\,
{\cal D}_a \,\frac{1}{\sqrt{U}}\, {\cal D}^a\, D \, , 
\label{3.11}
\end{equation}
${\cal D}_a$ denoting the covariant derivative with respect to the
metric $\gamma_{ab}$ (acting tangential to the hypersurfaces
$\Sigma_t$ of constant $t$). The scalar product 
\begin{equation}
\langle\chi_1|\chi_2\rangle\equiv 
\langle\chi_1|\chi_2\rangle_t = 
\int_{\Sigma_t} d\Sigma \, \sqrt{U} D^2 \,\chi_1^*\chi_2 \, ,
\label{3.13}
\end{equation}
with $d\Sigma$ the (scalar) hypersurface element on
$\Sigma_t$, 
serves to define the (formal) Hermitean adjoint $A^\dagger$ of
a linear operator $A$, and the complex conjugate of an operator
is defined by $A^* \chi = (A\chi^*)^*$. When $[A,S]=0$, the
operator $A$ acts tangential to $\Sigma_t$, i.e. when expressed
with respect to the coordinates $(t,\xi^a)$ it contains no
time-derivative. The time-derivative of an operator is defined by
$\dot{A}=[\partial_t,A]$. The operator $h$ satisfies
$h^\dagger=h^*=h$ and $[h,S]=0$. 

Consider now the following differential equations for an 
operator $H$
\begin{equation}
i \dot{H} = 2 h - 2 H - H^2 \, , 
\label{3.18}
\end{equation}
with the additional conditions 
\begin{equation}
[H,S] = 0 \qquad H^\dagger = H^*\, . 
\label{3.20}\\
\end{equation}
Whenever such an operator is given, any solution of the
Schr{\"o}dinger type equation 
\begin{equation}
i\,\partial_t \chi = H \chi 
\label{3.23}
\end{equation}
satisfies the wave equation (\ref{3.9}). The function $\chi$
thus corresponds to a solution $\psi$ of (\ref{2.2}). Denoting
the space of wave functions obtained in this way by 
$\H^+$ (and its complex conjugate by $\H^-$), the system of 
equations for $H$ implies that the space of wave functions
decomposes as a direct sum $\H=\H^+\oplus \H^-$ under the
relatively mild assumption that 
\begin{equation}
{\cal A} = 1 + \frac{1}{2} \Big(H + H^\dagger\Big)
\label{3.22}
\end{equation}
is a positive operator. Moreover, the Klein-Gordon scalar 
product $Q$ is positive (negative) definite on $\H^+$ ($\H^-$). 
When wave functions are rescaled once more as 
\begin{equation}
\eta = {\cal A}^{1/2}\,\chi \, , 
\label{3.35}
\end{equation}
we find
\begin{equation}
Q(\psi_1,\psi_2) = \langle\eta_1|\eta_2\rangle 
\label{3.36}
\end{equation}
for $\psi_{1,2}\in\H^+$ and 
\begin{equation}
Q(\psi_1,\psi_2) = - \langle\eta_1|\eta_2\rangle 
\label{3.37}
\end{equation}
for $\psi_{1,2}\in\H^-$, whereas 
$Q(\psi_1,\psi_2)=0$ if $\psi_1\in\H^+$ and $\psi_2\in\H^-$. 
Defining the operator 
\begin{equation}
{\cal K} = {\cal A}^{1/2} H {\cal A}^{-1/2} + 
i \,({\cal A}^{1/2})\,\dot{} \, {\cal A}^{-1/2}\, , 
\label{3.40}
\end{equation}
which is Hermitean, ${\cal K}^\dagger ={\cal K}$, 
and acts tangential to $\Sigma_t$, $[{\cal K},S]=0$, 
the Schr{\"o}dinger type equation (\ref{3.23}) becomes 
\begin{equation}
i\,\partial_t \eta ={\cal K}\eta 
\label{3.38}
\end{equation}
for $\psi\in\H^+$ (and an anologous equation for $\H^-$). 
Thus, with any choice $(S,D,H)$ there is associated a decomposition
of $\H$ and a unitary evolution of wave functions. 

This is not a very exciting result, but is becomes a  
bit more interesting when we ask how the decomposition {\it changes}
when the WKB-branch $(S,D)$ and the operator $H$ are varied 
by an infinitesimal amount $(\delta S,\delta D)$ --- compatible
with (\ref{2.5})--(\ref{3.6}) --- and $\delta H$ --- compatible
with (\ref{3.18})--(\ref{3.20}). 
It turns out that the decompositions of $\H$ defined in these
two branches actually {\it coincide} if 
\begin{equation}
\delta H = [H,\frac{\delta D}{D}] 
+i\,[H-h,\delta S] + [H,\delta S]\,(\partial_t + i H)\, . 
\label{4.10}
\end{equation}
Hence, if we manage to select a solution $H$ for any 
WKB-branch such that its variation $\delta H$ between
infinitesimally close neighbours satisfies the above equation, 
we expect a {\it unique} decomposition to be specified. 
(This should be true at least when any two WKB-branches
may be deformed into each other by a sequence of infinitesimally 
small steps). 
The big question is whether this can be done in a {\it natural}
way. 

The answer to this question is surprisingly simple, although at
a formal level. Rewrite the differential equation (\ref{3.18}) 
as 
\begin{equation}
H = h -\,\frac{1}{2}\, H^2-\,\frac{i}{2}\,\dot{H} 
\label{5.1}
\end{equation}
and solve it iteratively by inserting it into itself. 
This corresponds to the sequence 
\begin{eqnarray}
H_0 &=& 0\nonumber\\
H_{p+1} &=& h -\,\frac{1}{2}\, \,H_p^2 -\, 
        \frac{i}{2}\, \dot{H}_p
\label{5.5}
\end{eqnarray}
for non-negative integer $p$. One encounters a formal
expression whose first few terms read 
\begin{equation}
H = h -\,\frac{1}{2}\, h^2-\,\frac{i}{2}\,\dot{h} + 
\frac{1}{2}\, h^3 +\frac{i}{2}\,\{h,\dot{h}\}-\, 
\frac{1}{4}\, \ddot{h} + \dots
\label{5.2}
\end{equation} 
When appropriately keeping track of the various terms appearing in this
procedure (for details see Ref. \cite{FE10}), 
we can reformulate $H$ as a formal series, called
the ''iterative solution''. The amazing thing is that it
formally satisfies the condition (\ref{4.10}) guaranteeing the
decomposition to be unique. However, it is not quite clear in
which models this series will actually converge (or, on which 
wave functions it will converge). 

In case of convergence, we have specified a solution $H$ in 
any WKB-branch without ever having performed a choice!
Due to the appearance of a series containing arbitrarily
high time-derivatives of $h$, we expect $H$ to rely on
{\it global} properties of the model, possibly connected with
{\it analyticity} issues. 
In Ref. \cite{FE10}, these manipulations have been presented in
detail, and there the question is raised whether the structure
encountered is related to quantization methods that start
from global features, such as the scalar product
$q$ from (\ref{q}), and some examples for the iterative
solution are given. 

\section{EXAMPLE: FLAT KLEIN-GORDON EQUATION}

A first orientation about what can be achieved by this 
approach is to apply it to the Klein-Gordon equation in 
flat space ($ds^2=-dt^2+d\vec{x}^2$ and $U=m^2$, which we set
equal to $1$) as a toy model. In this case $h=-\frac{1}{2}\triangle$,
hence $\dot{h}=0$, and the iterative solution (\ref{5.2}) 
becomes the closed expression 
\begin{equation}
H= \sqrt{1+2h}-1\,. 
\label{sqrt}
\end{equation}
Note that it is well-defined although (\ref{5.2}) does not
converge on all wave functions. Similar features may be expected
in more general models as well, so that it is not at all obvious
how to give the formal solution a mathematically well-defined 
meaning. The spaces $\H^\pm$ in our example are 
identical with the standard positive/negative frequency
subspaces of $\H$. 

\section{DISCUSSION}

Given that the procedure described in Section 2 goes through in 
a particular model without symmetries, 
the question arises what structure one has touched upon. 
Due to the existence of a preferred decomposition, there is 
a natural positive definite scalar product on the space of wave 
functions, provided by 
$(\H,Q_{{}_{\rm phys}})\equiv(\H^+,Q)\oplus (\H^-,-Q)$. 
In the case of the Klein-Gordon equation in flat space, this
just amounts to reverse the sign of $Q$ in the 
negative frequency sector. On the other hand, the refined algebraic
quantization program arrives at a Hilbert space 
$({\cal H}_{{}_{\rm phys}}, \langle\,|\,\rangle_{{}_{\rm phys}})$
which, for the flat Klein-Gordon equation, agrees with
our construction \cite{flat}. Hence, it is 
natural to ask whether
this relation carries over to more general cases. 
In Ref. \cite{FE10}, an independent argument is given that the mere
coexistence of the indefinite Klein-Gordon scalar product $Q$ and the
physical inner product $\langle\,|\,\rangle_{{}_{\rm phys}}$ 
make the existence of a preferred decomposition very likely. 
 
Unitarity shows up in our approach only in the context of
WKB-branches. This resembles the way tensor components show
up in coordinate systems. Consequently, there is not ''the'' unitary
evolution, although one has the feeling to deal with 
a fundamental structure. Maybe the expression 
''covariance with respect to WKB-branches'' is appropriate 
to describe this point of view. 

In a semiclassical context, the operator $h$ is 
expected to be ''small'', so that the first few terms in 
(\ref{5.2}) 
may be regarded as numerically approximating 
the exact solution $H$. Using physical units, in a typical 
quantum cosmological setting we have 
$H\approx h\,+$ Planck scale corrections. 
The first few terms of the operator (\ref{3.22}) defining the
redefinition (\ref{3.35}) are given by 
\begin{equation}
{\cal A} =  1 + h -\,\frac{1}{2}\,h^2 + 
\frac{1}{2}\,h^3-\,\frac{1}{4}\, 
\ddot{h} + \dots, 
\label{5.12}
\end{equation}
which might in a concrete model serve to justify the assumption that 
$\cal A$ is positive. 
The Hermitan operator (\ref{3.40}) defining the unitary evolution
(\ref{3.38}) is given by 
\begin{equation}
{\cal K}=h -\,\frac{1}{2}\,h^2 + 
\frac{1}{2}\,h^3 -\,\frac{i}{8}\, 
[h,\dot{h}] - \,\frac{1}{4}\,\ddot{h}+ \dots
\label{5.15}
\end{equation}

It is not claimed here that 
(\ref{3.38}) is ''the'' Schr{\"o}dinger equation
with $t$ as the physically experienced time, but our
result opens the perspective of deriving conventional
quantum physics within the context of the familiar
mathematical framework of quantum mechanics. 

\section{ACKNOWLEDGMENTS}

Work supported by the Austrian Academy of Sciences
         in the framework of the ''Austrian Programme for
         Advanced Research and Technology''.

\end{document}